# Reconfigurable Inspection in Manufacturing: State of the Art and Taxonomy

Harshit Gupta
*Department of Mechanical Engineering*
*Delhi technological Univeristy*
Delhi, India
harshitgupta_me20a9_36@dtu.ac.in

Ashok Kumar Madan
*Department of Mechanical Engineering*
*Delhi Technological University*
Delhi, India
ashokmadan@dce.ac.in

*Abstract*—This article provides an overview of the evolution of the product quality and measurement inspection procedure with emphasis on the Reconfigurable Inspection System and Machine. The major components of a reconfigurable manufacturing system have been examined, and the evolution of manufacturing processes has been briefly discussed. Different Reconfigurable Inspection Machines (RIMs) and their arrangement in an assembly line as an inspection system have been carefully studied and the modern inspection system equipped in RMS has been compared to the traditional techniques commonly used in inspection of product quality. A survey of evolving inspection techniques is offered from the standpoint of technological challenges and advancement affecting manufacturing over time. As per authors' knowledge, the review on Reconfigurable Inspection in Manufacturing and taxonomy of reconfigurable inspection systems is rare. Considering the studies done in this domain, there is still resourceful taxonomy for this paradigm. Therefore, different types of inspection procedures have been discussed, their features and applications have been compared to arrive at the taxonomy of the RIS based on the understanding of the nature of a RIS after a critical review.

## I. Introduction

Product quality is a crucial factor for producers in a modern production environment. The outcomes of poor product quality can be experienced in many varied fields, including resource waste, increased environmental impact, and direct financial losses. Hence, quality management is of utmost essential to keep up with financial and operational performances [1]. Subpar products and unsatisfied clients are some negative social impacts that poor quality products can have [2]. The sustainability of production systems and processes must also be a top priority for the modern manufacturing sector and industries must balance social, environmental and economic considerations [3]. Product quality inspection is the foundation of Sustainable Manufacturing as it focuses on implementing Zero Defect Manufacturing [16].

Manufacturing companies must enhance their standards for the management of products and processes due to significant changes in the modern industrial environment. The changeover from conventional manufacturing systems to the Next Generation Manufacturing Systems (NGMSs) is being fueled in part by factors including greater flexibility and customization, shifting market demand, adaptive batches, high-quality items and short product life cycles [31]. Manufacturing businesses must adopt a new type of arrangement that is highly responsive to global markets, a type whose capacity can be adjusted to changes in demand and is built to be elevated with new process technology required to accommodate complex product specifications. Even the Flexible Manufacturing Systems (FMSs) now lack these qualities. The pillars of this new manufacturing paradigm are efficient reconfigurable systems, whose components are reconfigurable machines (reconfigurable tool, inspection, assembly).

As with mass production and lean manufacturing, the manufacturing paradigm known as "Reconfigurable Manufacturing" will have a significant impact [32]. Reconfigurable Manufacturing Systems (RMS) act as a possible solution to cater to the demands and meet the responsiveness in the market. According to [4], "RMS is designed at the outset for rapid change in structure, as well as in hardware and software components, in order to quickly adjust production capacity and functionality within a part family in response to sudden changes in the market or regulatory requirements". Reconfigurable Inspection systems (RIS) on the other hand, is a type of RMS that uses reconfigurable tools for the purpose of inspection of product quality so that both market demands and customer expectations can be met. Modular inspection is necessary to be based on the Total Quality Management Principle. [27]

This study is structured as: Section II briefly describes the adopted review methodology. Section III discusses the evolution of manufacturing systems over time and the emergence of RMS with its components. Section IV showcases the literature review on RIS based on which a taxonomy is proposed which provides novelty to the paper. Section V entails the findings from the review and conclusion with scope to some future research work in the domain.

## II. Review Methodology

A systematic approach of pointing out, assessing and summarizing the completed and recorded works created by researchers and scholars on a certain topic is known as a literature review. The following steps in a chronological order make up the process for the literature review: Problem formulation, literature search, literature collection, quality assessment, analysis and synthesis, interpretation, and presentation of the findings. For this study, we obtained information and knowledge from Web of Science, Scopus and

IEEE Explorer— three most frequently used search engines. Also, some data and information were acquired from papers presented at conferences like the well-known IFAC conference on automation. The search has been performed using the search string that comprised of five groups of keywords:

•Reconfigurable Manufacturing: ''reconfigurable'' OR ''modular*''

•Manufacturing industry: ''manufactur*'' OR ''production''

•Automation: ''automat*'' OR ''sensor''

•Quality inspection: ''inspect*'' OR ''measure'' OR ''quality''

•Quality inspection type: ''inline'' OR "offline" OR "online"

The article search was focused within the field of engineering and manufacturing and written in English.

### III. EVOLUTION OF MANUFACTURING SYSTEMS

The evolution of a manufacturing system depends on the nature, need and demand of the product it produces. There has been a massive shift from mass production as in the case of Dedicated Manufacturing Lines (DML) to mass customization (MC) and personalization (FMS) to make processes customer-centric and responsive [33]. This has resulted in a massive change in quality, productivity and responsiveness in manufacturing. The development of dedicated manufacturing lines as a component of the Dedicated Manufacturing System (DMS), which manufactured vehicle engines, transmissions, and other key parts, was the only thing that allowed for mass manufacturing. Such specialized production lines manufacture a single item type at a very high production rate, and they are highly profitable when there is a large demand for that particular part. Until the middle of the 1990s, these dedicated transfer lines generated huge quantities of items and were recognized as the most profitable systems. [19]

The development of FMS in the early 1980s was made possible by the discovery of NC (numerically controlled) and then CNC (Computer Numerically Controlled) in the 1970s [5]. In 1981, Stecke and Solberg defined the operating principles of an FMS having nine machines connected by an autonomous material handling mechanism for a specific job shop [6, 7]. Yet, it took nearly 20 years for FMSs to enter the sector of transportation powertrain because of the high initial investment cost. Productivity, quality, and flexibility were the strategic objectives of manufacturing businesses implementing FMS. Midway through the 1990s, increased globalization and international competitiveness made it evident that FMS only offered a partial solution from the economic perspective. FMS's conventional structure employed in the industry made it easier to modify the goods that were made, but it produced a relatively slow pace of production and lacked the volume flexibility needed to react to sudden changes in demand brought on by international competition. Manufacturing system designs should be able to meet the strategic objectives of the company and help the company to sustain to satisfy demands [8].

The physics of the system cannot be changed in an economically feasible manner to scale a system using classical FMS. The solution to this was the adaptation of RMS. The University of Michigan submitted a request for proposals to the National Science Foundation (NSF) in 1995 for the creation of a research center on Reconfigurable Manufacturing Systems (RMS). According to the report, an RMS can rapidly adapt its production capacity and functionality in response to changing conditions by rearranging or reconfiguring or replacing its hardware and software components unlike FMS that reconfigures only the hardware component or unlike DMS that doesn't show any replacement or rearrangement of the components [32]. This new System of manufacturing would be the best of both worlds i.e., the DMS and the FMS and it would not only allow personalization and responsiveness to market demands but would also help ramp up productions. [19].

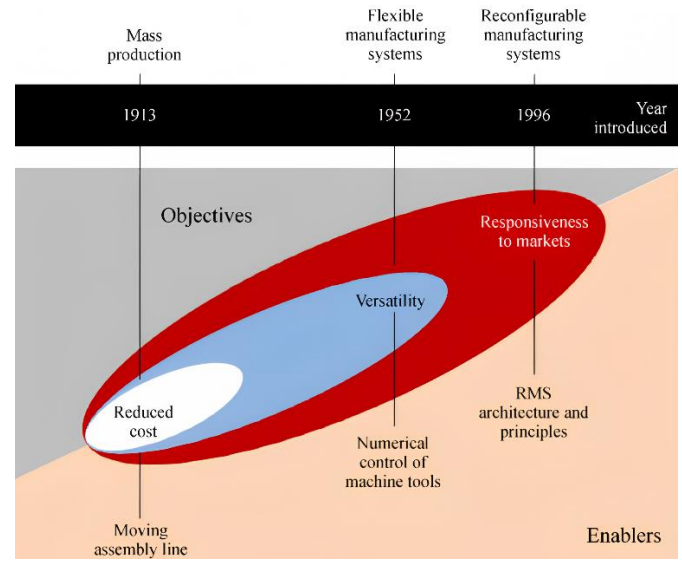

Fig. 1. A paradigm shift in the Manufacturing Systems [23].

#### A. Reconfigurable Machines and Its Types

Reconfigurable machines (RMs), which impart the systems the flexibility to tailor and react rapidly to evolving and changing market conditions, are the main building blocks of RMSs. These can carry out a variety of tasks because of its modular structure, which is made up of physical parts that are easily swapped out and/or rearranged. This enables an RM to be modified to produce a variety of goods, making it able to adjust to a novel good or fill an advance need [21]. According to [9], RMs are broadly categorized into the following four types (shown in Figure 2):

#### B. Reconfigurable Machine Tools

Reconfigurable machine tools (RMTs) are physical modules (such as spindle heads, machining tools, turrets) that are used in machining centers and each of which is capable of carrying out one or more production jobs. These modules can be arranged inside a machine in a variety of ways, offering many combinations [21]. The RMT's design is often concentrated on a single part family, and it should be able to quickly adapt to alterations in its operations or structure to manufacture different portions related to that specific part family. The first ever RMT patent was granted in 1999 [10]. Figure 3 depicts an arch-type RMT constructed in 2002 to drill and mill perpendicular to the surface when working on slope surfaces. This RMT may be adjusted to five possible spindle axis angular settings in steps of

15° between -15° and 45°exhibiting its modularity, and changing from one angle to another takes less than two minutes. Engine blocks were milled and drilled using it at angles of 30° or 45° [18]. A new design methodology for identifying the best reconfigurable machine tool, taking into account both machine configurations and the reconfiguration process, is shown and described in [30].

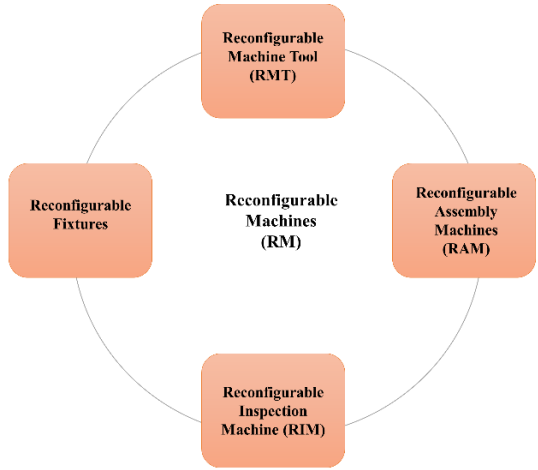

Fig. 2. Components of an RMS system.

### C. Reconfigurable Assembly Machines

The assembly and transfer lines make use of Reconfigurable Assembly Machines (RAM). These machines have the ability to be modular and hence can be modified or altered to assemble parts to form products having certain common characteristics [21]. The automotive heat exchanger proves to be a valuable example for a reconfigurable assembly machine [35]. The enormous number of potential modules that might be combined to create a complex product presents one of the main obstacles to mass individualization. Because of this, an assembly and transfer line system should have the ability to generate a wide variety of models and variations of the same part family [18].

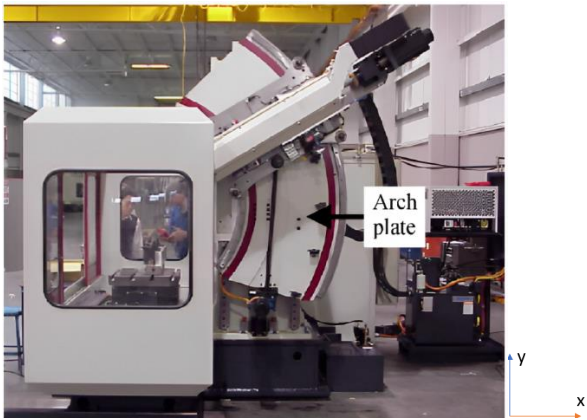

Fig. 3. Arch type RMT [18].

### D. Reconfigurable Fixtures

Substantial or intricate components, such as an engine cylinder heads are machined with the help of reconfigurable fixtures. The part is mounted on custom fixtures during the process so that various machining operations, including cutting and welding, can be carried out on its surface [21].

### E. Reconfigurable Inspection Machine

Inspection and measurement of the machined parts that are currently being worked on is carried out by a reconfigurable inspection machine (RIM). They are mostly made up of electro-optical sensors. Depending on the component to be measured, the number and placement of these sensors can fluctuate [21]. A family of in-process inspection devices known as reconfigurable inspection machine can be adjusted to fit the geometrical measurements and quality of the inspected part. RIM received a ground breaking patent in 2003. [11].

## IV. LITERATURE REVIEW

### A. Reconfigurable Inspection

Although being built to incorporate responsiveness and rapid changes, when an RMS is reconfigured, there are probably going to be a lot of quality issues that need to be addressed. For instance, there is always potential for misalignments and resultant dimensional mistakes with each module that is moved about in the system. These high-quality issues hinder the ramping up process after reconfiguration and result in system downtime. The ramping up capacity and performance of reconfigurable manufacturing systems, as well as all large volume production systems generally, depend on the quick elimination of product quality problems [25].

In the traditional manufacturing (FMS and DML) and inspection techniques, touch type of inspection doesn't take into consideration time over repeatability and accuracy. This category encompasses the Coordinate Measurement Machine (CMM), force sensors and touch probes for inspection purposes. These days, producers use the industry standard CMMs and other highly adaptable inspection equipment to lower the cost of adjusting to changes in manufacturing. These machines' inspection rates, however, are substantially slower than their output rates. Due to this, manufacturers are compelled to periodically sample items off-line in order to check for quality. As a result, quality issues are frequently not discovered until a sizable number of unacceptable parts have been manufactured which is quite disruptive and expensive [25]. The "virtual ball" approach was created and used to contrast the results obtained by RIM measurement with those of a CMM. [37]. It offers interpretations of measurements of non-contact laser as if taken by a CMM touch-probe. In [22], repeatability data from the RIM are shown, along with a comparison of measurements made using the RIM and a CMM.

Advanced manufacturing systems must be able to determine a manufacturing process's present status automatically. Processing fault detection can guarantee that mistakes are discovered quickly so that the machine function and operation can be adjusted, decreasing the production of unsatisfactory products [12]. The reconfigurable inspection system (RIS), which is made up of several RIMs, is used in an RMS for product quality assessment. A design technique for RIS configuration is necessary for dynamic management in RMS-based reconfiguration and for an effective viewing of the production process. A tailored detection system that is sensitive to process mistakes and capable of gathering sufficient information to facilitate the determination of the underlying cause of error is provided by a RIS configuration design.

To address the above requirements, [24] suggests a technique for investigating a design strategy that guarantees the inspection system can do thorough and fast detection by utilizing a small number of inspection stations and machines. An inspection station follows each operational station in an RMS. This strategy, though, is expensive and may not be practical. To oversee the production process, it is crucial to choose a small but necessary number of inspection stations [24]. Exploring sensor embedded technologies used in the process of product quality inspection is equally vital as designing a Reconfigurable Inspection System with a minimal and optimum configuration. An RMS includes delivery and return systems, operating stations, inspection stations, and inspection stations as described in figure 4. The authors contend that there is a complicated flow of information throughout the production process as a result of the interrelatedness of the effects of the various stations on product quality. To describe the information flow of the process of manufacturing, the use of Stream of Variation (SoV) theory [13] has been proposed by the authors to model the propagation and accumulation of faults and errors in the manufacturing process, as shown in figure 5.

In figure 5 shown below, variables X(k), U(k), and Y(k) denote the measurement vector of the kth station, the accumulated deviation of the station 'k', and the processing deviation supplied by the station 'k', respectively. Moreover, processing and measurement noise, respectively, is ξ(k) and η(k).

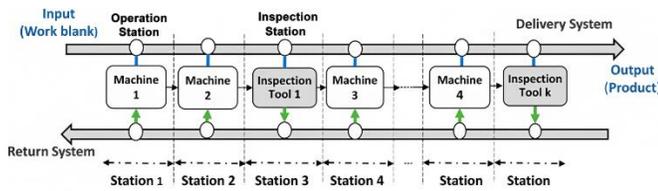

Fig. 4. RMS Manufacturing Process [24].

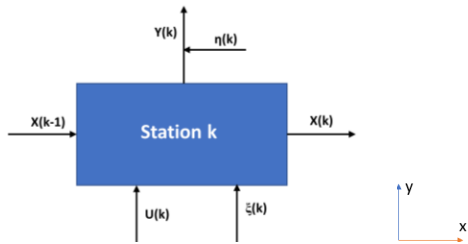

Fig. 5. RIS Model according to SoV [24].

[24] Sensor fusion is neglected by the authors while devising the optimal configuration between the various components. However, the quality of processing of the headstock which directly impacts the accuracy, life and performance has been studied and verified by the proposed method where cost and diagnosability were verified for each configuration.

Sensor Fusion through the usage of non-contact laser and optical sensors however has been studied by [23]. A sample RIM can be shown in Figure 6, which is made up of a conveyor that moves the part along a precise axis of motion inside of a variety of electro optical equipment, like line scanning digital cameras and laser-based sensors. The positioning and quantity of sensors in the reconfigurable machine type can be altered to meet the shape of the part already inspected, depending on the part that is being measured. The RIM shown in Fig. 6 was set up for engine cylinder head measurements. The position of sensors is such that on one side of the component, there are three laser sensors; on the other, there are two laser sensors and a precise computer vision system. The authors also used the SoV methodology for identification of root causes of product quality errors as the product passed on from one station to another.

SoV method may be used extensively but is not the only algorithm to be applied to an RIS. An example of a Reconfigurable Inspection Machine was described and tested on the product family of torches by the proposed quality management method [27]. To determine the physical parameters of the product, a first reference part was required. A reconfigurable pallet was created to handle the family of cylindrical torch parts. This pallet may be adjusted for the cylinder and head unit's length and diameter. Drilled holes were used to achieve the pallet structure's discrete location. The revolution of the spring-loaded prongs while ensuring safety of the torch made it easier to make further the adjustments. Due to the range of measurements that can be taken, vision was chosen as the inspection method. The method includes inspecting the product's length, color, labelling, and feature detection. Unlike the authors at [24] and [25] who used the SoV (stream of Variation) algorithm, the authors used the greedy-first search algorithm to calculate the most optimal and efficient inspection route but with respect to time.

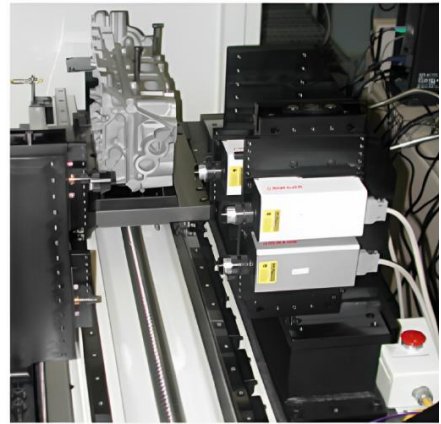

Fig. 6. RIM system [23].

[26] concentrates on introducing the RAVIS (Reconfigurable Automated Vision Inspection System) into the CIM (Computer Integrated Manufacturing) cell to increase agility to the inspection capacity of an RMS. RAVIS, is a product quality control device to assess the caliber of produced components from a product family. It uses a laser-based method to gather information through the creation of 3D images. Using a reverse engineering method, quality assurance testing and product development can be carried out with the images created. The Reconfigurable Automatic Vision Inspection Systems (RAVIS) geometrical captures details using a design framework

for reverse engineering. After that, procedures processed the data to precisely characterize the product quality.

[36] defines a modular visual inspection method for detection for non-planar features of additively manufactured parts. The system has two modules: one handles picture acquisition, the other handles image analysis. A module that handles photos is needed in the interim, and it must be given images from the module handling acquisition so that the analysis module can access images of product areas independent of the method of acquisition. Its modularity is intended to offer a system that can be quickly and readily expanded to support new products.

[38] presents and discusses a visual inspection system designed for a robotic work cell in the context of an automotive light (headlight) assembly example. A binary quality control task has been carried out to check for damaged or incorrectly placed parts during the assembly procedure. A 2D camera has been used to capture images and detect errors. The framework has been prepared using ROS and OpenCV libraries for the purpose of Simulation, verification and analysis of the data collected from the various sensor nodes. The image processing and computation technique used by the authors has been shown in Figure 7.

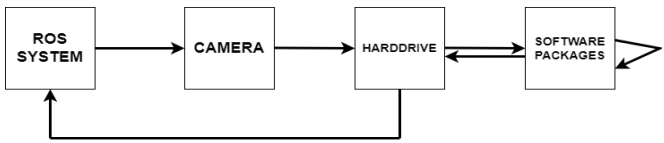

Fig. 7. Image Processing Pipeline using ROS and OpenCV [38].

*B. Mode of Reconfigurable Inspection*

The two common paradigms that the authors of [29] classify the inspection process into are "offline" and "online" inspection. In an offline inspection procedure, while the units are inspected offline i.e., separately from the assembly or production belt once the production process is finished, they are inspected while the units are being manufactured in an online inspection procedure. Conventional online inspection procedures are often more cost-efficient and productive than equivalent offline procedures [34]. Online inspections, however, are not always practical because of the nature of activity or the available time [14]. Offline inspections can be carried out when the product is semi-finished at different production phases or when it is fully finished at the end of the assembly line [15]. The lead time is another disadvantage in an offline inspection process as it is more than the online inspection process as the part needs to be removed and added back at various stages of manufacturing and assembly process.

Based on the above classification, a reconfigurable inspection system should be categorized as a type of "online" inspection procedure wherein the product after processing through the various operation workstations is inspected at the inspection stations without going "off" the assembly line. Therefore, there are separate workstations for implementing the manufacturing process and for implementing inspection procedures as shown in figure 4 and the part is either rejected or sent to the previous workstation for reworking.

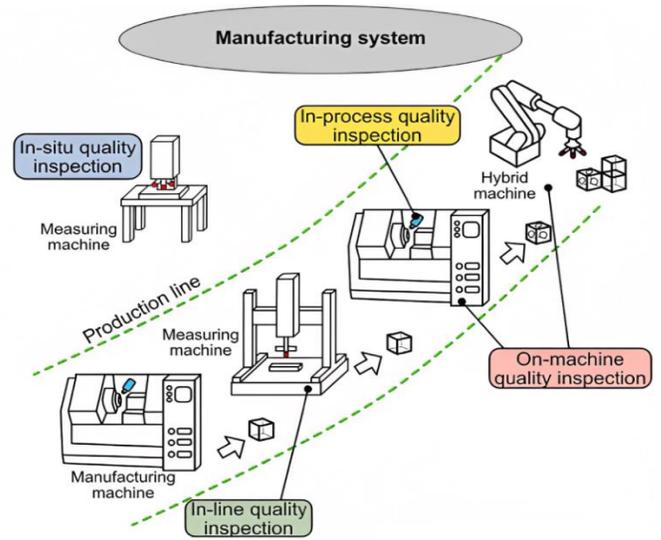

Fig. 8. Schemes for quality inspection [28].

In the industrial sector, the word "in-line" is frequently used to specify the time and location of measurements, i.e., on the line of production or during operations that adds value to the product features and functions [28]. There have been several "in-line" sub-terms that have been given through the years. These terms that are used synonymously are in-situ quality inspection, on-machine quality inspection and in-process quality inspection. Figure 8 shows the scheme of condition for quality inspection as discussed by [17]. The inspection station is often the last station in a standard manufacturing chain. End-line detection is a well-liked strategy in the sector and is thought to be cost-effective. However, only some faults can be recognized. Thus, based on the representation and classification of manufacturing systems shown in figure 8 and the categorization discussed above, the in-process quality inspection scheme is similar to that of an inspection station as described in a Reconfigurable Inspection System and Machine.

*C. Taxonomy of Reconfigurable Inspection Systems in Manufacturing*

The literature study carried out by the authors in the above section of the various types of Reconfigurable Inspection Systems and Machinery and their varying applications leads to the formulation of a taxonomy of RIS in Manufacturing Systems. Table 1 describes an initial comprehensive overview of the structure, behavior and function of the various inspection systems discussed in literature. Table 1 is a summary of various characteristics of the inspection system such as Method of Inspection which provides insights into the type of sensors used (the physical dimension of the system), the Algorithm used provides the necessary computation insight and the Application of the system that highlights the function of the system to draw comparison and find relationships between the function, behavior and structure of such inspection systems in an RMS. The table also provides an initial start to the construct of the taxonomy proposed later in this section.

TABLE I. CHARACTERISTICS OF VARIOUS RISS

| S.no. | Ref. | Method of Inspection | Algorithm | Application |
|---|---|---|---|---|
| 1 | [24] | - | SoV | Processing quality of headstock |
| 2 | [23] | Laser and Computer Vision | SoV | Engine cylinder head measurements |
| 3 | [27] | Vision | Greedy search algorithm | Length, Color, labelling, and feature detection of torches |
| 4 | [26] | Laser and Computer Vision | Reverse Engineering of 3D image | Integrate and enhance CIM operation to easily characterize product quality. |
| 5 | [36] | Vision | SWARTLAM | Verification of Foil alignment |
| 6 | [38] | Vision | OpenCV (Hough transform) | Quality control of damaged parts and misalignment in automotive headlights |

We have researched a number of resources where reconfigurable inspection is discussed and shown in light of the paper's goal. When we were creating our taxonomy, we discovered the reconfigurable inspection footprint in a variety of research and application types. Through a survey, sanitization process, and an explanation of works in the field, the aim is to conceptualise it. Utilizing both, literature reviews and systematic research techniques of various literature resources, a taxonomy of the Reconfigurable Inspection is developed into five categories. The proposed taxonomy is regarded as a first step that could potentially lead to further development. The proposed taxonomy opens up possibilities for a deeper comprehension of this new manufacturing ecosystem. For generating the proposed taxonomy, we studied and analyzed recent articles, literature on various databases. With the help of five categories, related fields, and disciplines, the key elements of the proposed Reconfigurable Inspection taxonomy are described:

i. Components: A reconfigurable inspection system and a machine would consist of a physical component comprising of non-contact sensors and fixtures for positioning and holding the workpiece and a cyber component to carry out computations by various algorithms on data collected by the sensors in place. For example, variety of cameras and laser sensors are used for capturing of images and collection of data points which are then analyzed using suitable algorithms such as SOV, greedy search, Hough Transform (OpenCV) depending on the function of the inspection system employed.

ii. Measurement Type: The process of inspection is carried out to measure the quality or the features/dimensions of the product being processed in the transfer line. In manufacturing assembly lines, regular inspections are also carried out to check for alignment of the products being assembled. A total of six different applications of inspections have been discussed on the basis of which common characteristics of measurement of features, quality and misalignment have been found. These quality control measurements also form the pillars of quality management and inspection in an industrial setup.

iii. Design Features: The RIM is designed for customized flexibility because it can measure a variety of product features ranging from color, labelling, dimensional measurements to overall quality. It is also made to be quickly and simply convertible by adding sensors as needed and moving existing sensors to different locations as needed for inspection of various parts or various features of the same part. The RIM is made to be scalable, i.e., to permit mounting of various probes at various starting points. Scalability makes it possible to measure a larger number of features effectively. Modularity of an RIS allows it to combine different modules involved in the inspection procedure together and work in sync to carry out the common objective of maintaining consistent quality.

iv. Inspection Methodology: A reconfigurable inspection makes use of laser or vision (or combined) technology to inspect the part and make suitable decision to send it for reworking or discard it. Based on the study of relevant literature, laser and vision are found to be the most common types of methods applied for non-destructive, effective and efficient inspection of part dimensions, features, quality and misalignment.

v. Nature of Inspection: Based on our discussion in Section 5.B, the mode of inspection of such system can be regarded as "on-line" and "in-line" as the product inspection takes place "on" the transfer line and "during" the manufacturing phase. The two modes are compared against "off-line" and "end-of-line" that are usually found in conventional manufacturing systems as discussed in section 4.A. Such type of inspection makes the process faster as it allows for instant reworking of the part thus saving cost and time to market as the part doesn't have to be removed and put on successively which further prevents any damage that might occur as the part is never removed from the transfer line until the quality is satisfactory.

The taxonomy has been shown in Figure 9. The various characteristics of the reconfigurable inspection branch out which further branch out into their respective components. RIS works collectively by combining the components and using a particular method of inspection by following specific design features and nature of inspection to achieve a particular objective of carrying out a specific type of measurement (Measurement Type).

Each category aims to present a different critical dimension of the system, thus, the proposed taxonomy makes it easier to understand RIS and its characteristics. Such a taxonomy is vital to help researchers and manufacturers know the key aspects and features of such inspection systems if they plan to implement such systems for their facilities. The taxonomy is also leveraged to identify open problems that can lead to new research areas within this domain and can be further expanded on basis of common characterstics among RISs.

V. CONCLUSION AND FUTURE SCOPE

Reconfigurable manufacturing systems combine the best of both Dedicated manufacturing systems and flexible manufacturing systems to guarantee customer satisfaction, higher production, agility and higher quality products leading to ensuring higher levels of quality and measurement inspection. With the evolution of manufacturing systems, there has been a massive shift and upgrade in the inspection procedures carried out to test product quality as discussed in the literature review. Different reconfigurable inspection systems [23,24,25,27,36,38] that provide the flexibility to diagnose the product quality and measurements have been discussed.

It was observed that compromise must be made between the cost and the diagnosing capability of the complete system to provide the optimal arrangement of the operation and inspection workstations in an RIS which can be done by statistical methods such as the Stream of Variation as discussed by [24]. The SoV method was also used by [25] to accumulate the root cause of errors while inspecting an engine block. However, the authors at [27] used a greedy-first search algorithm to calculate the most efficient inspection route for a product family of torches using time as the performance parameter. This indicates that time to market and the cost of production of a particular family of products in a reconfigurable environment play a key role apart from providing high quality products. From the mentioned studies and classification of inspection procedures provided by the authors as discussed, it can be concluded that inspection procedure in a reconfigurable inspection system can be regarded as an online inspection process and the methodology of inspection in such modular environment with modularity in machine functions and position, is similar to the in-process quality inspection scheme.

an overview of the Function, Structure and Behavior of the system that can provide valuable insights into the overall performance of such systems. Future studies can expand this taxonomy or revamp it for the user's best interests and further contribution to the research domain.

As part of future work in this research domain, researchers can also look for various other algorithms apart from the ones already discussed in the paper to examine the inspection methodology in an RIS and also compare them to the ones described in this review to arrive at a suitable conclusion of the best fit algorithm based on varying scenarios and parameters. Future research work in RIS may also comprise of using some other non-contact sensors to gather critical quality and dimensional information about the product during its operation in an RMS. The study of variety of fixtures in an RMT during the inspection process and accounting for possible misalignment between various sensors and workpiece holding devices and its impact on the overall readings and quality measurement is another work that researchers can take up and come out with significant contributions.

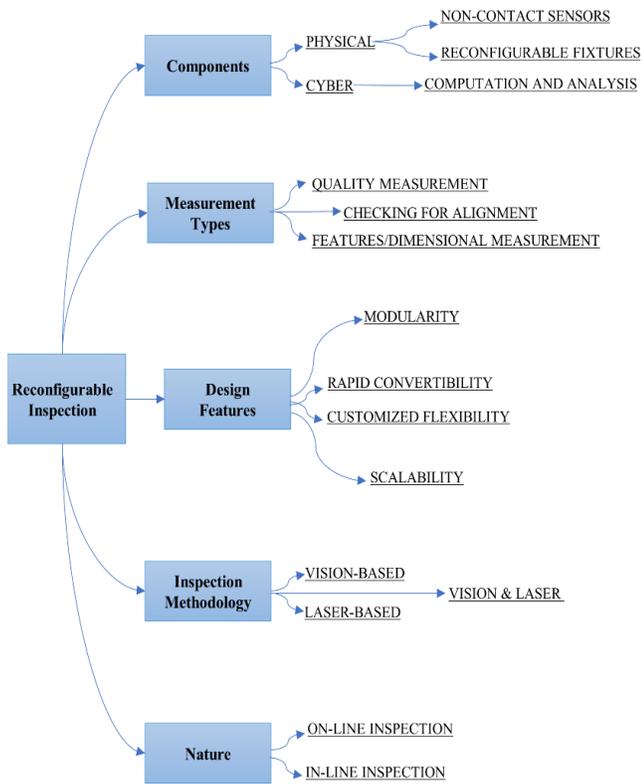

Fig. 9. Reconfigurable Inspection Taxonomy.

Based on the critical review of papers, a taxonomy is proposed that will help formulate and conduct future research and prototype products related to Reconfigurable Inspection systems in a more structured way and method. The resulting taxonomy is of great importance to study and interpret the nature and functioning of an RIS. Taxonomy of such systems provide

REFERENCES

[1] Kumar P, Maiti J, Gunasekaran A., 2018, "Impact of quality management systems on firm performance." Int J Qual Reliab Manag

[2] Jun J-h, Chang T-W, Jun S., 2020, "Quality prediction and yield improvement in process manufacturing based on data analytics." Processes;8(9):1068.

[3] Lu Y, Xu X, Wang L., 2020, "Smart manufacturing process and system automation– a critical review of the standards and envisioned scenarios." J Manuf Syst; 56:312–25.

[4] Koren, Y., Heisel, U., Jovane, F., Moriwaki, T., Pritschow, G., Ulsoy, G., and Van Brussel, H., 1999, 'Reconfigurable Manufacturing Systems', CIRP Annals, 48, (2), pp. 527-540

[5] Koren Y., 1983, "Computer control of manufacturing systems." McGraw-Hill

[6] Stecke KE., 1983, "Formulation and solution of nonlinear integer production planning problems for flexible manufacturing systems." Management Science; 29(3):273–88.

[7] Stecke KE, Solberg J.,1981, "Loading and control policies for a flexible manufacturing system." International Journal of Production Research,19(5):481–90.

[8] Cochran DS, Arinez JF, Duda JW, Linck J., 2001–2002 "A decomposition approach for manufacturing system design." Journal of Manufacturing Systems; 20(6):371–89.

[9] Koren, Y. 2010. The Global Manufacturing Revolution. Wiley

[10] Koren Y, Kota S. US Patent 5943750, 1999-08-31

[11] Koren Y, Katz R. US Patent 6567162, 2003-12-24

[12] Koren, Y. 2013. "The Rapid Responsiveness of RMS." International Journal of Production Research 51 (23-24): 6817–6827.

[13] Ding, Y., D. Ceglarek, and J. Shi. 2000. "Modeling and Diagnosis of Multistage Manufacturing Processes: Part I – State Space Model." In Proceedings of the 2000 Japan/USA Symposium on Flexible Automation, Ann Arbor: Michigan, pp. 23–26.

[14] Raz, T., Y. T. Herer, and A. Grosfeld-Nir. 2000."Economic Optimization of off-Line Inspection" IIE Transactions32(3):205–217.

[15] Tzimerman, A., and Y. T. Herer. 2009. "Off-line Inspections Under Inspection Errors." IIE Transactions 41 (7): 626–641.

[16] Psarommatis F, May G, Dreyfus P-A, Kiritsis D., 2020 "Zero defect manufacturing: state of-the-art review, shortcomings and future directions in research." Int J Prod Res;58(1):1–17.

[17] Gao W, Haitjema H, Fang F, Leach R, Cheung C, Savio E, 2019, "Onmachine and in-process surface metrology for precision manufacturing." CIRP Ann;68(2):843–66.

[18] Koren, Y., Gu, X. and Guo, W., 2017 "Reconfigurable Manufacturing Systems: Principles, design, and future trends," Frontiers of Mechanical Engineering, 13(2), pp. 121–136.

[19] Koren, Y. and Shpitalni, M.,2010, "Design of reconfigurable manufacturing systems" Journal of Manufacturing Systems,29(4).

[20] Wang, W. and Koren, Y., 2012, "Scalability planning for Reconfigurable Manufacturing Systems," Journal of Manufacturing Systems, 31(2), pp. 83–91.

[21] Yelles-Chaouche, A.R., 2020, "Reconfigurable manufacturing systems from an optimization perspective: A focused review of literature," International Journal of Production Research, 59(21), pp. 6400–6418.

[22] Katz, R., 2006, "Design principles of reconfigurable machines," The International Journal of Advanced Manufacturing Technology, 34(5-6), pp. 430–439.

[23] Koren, Y., Gu, X. and Guo, W., 2017, "Reconfigurable Manufacturing Systems: Principles, design, and future trends," Frontiers of Mechanical Engineering, 13(2), pp. 121–136.

[24] Xinwen Shang, Jelena Milisavljevic-Syed, Sihan Huang, Guoxin Wang, Janet K. Allen & Farrokh Mistree , 2020, "A key featurebased method for the configuration design of a reconfigurable inspection system", International Journal of Production Research.

[25] Barhak, J., 2005, "Integration of reconfigurable inspection with stream of variations methodology," Int Journal of Machine Tools and Manufacture, 45(4-5), pp. 407–419.

[26] Xing, B., 2006, "Reconfigurable manufacturing system for agile manufacturing," 12th IFAC/IFIP/IFORS/IEEE/IMS Symposium Information Control Problems in Manufacturing.

[27] Shaniel Davrajh Glen Bright, 2013,"Advanced quality management system for product families in mass customization and reconfigurable manufacturing", Assembly Automation, Vol. 33 Iss 2 pp. 127 - 138

[28] Azamfirei, V., Psarommatis, F. and Lagrosen, Y., 2023, "Application of automation for in-line quality inspection, a zero-defect manufacturing approach" Journal of Manufacturing Systems,67:1–22.

[29] Gianfranco Genta, Maurizio Galetto & Fiorenzo Franceschini (2020): Inspection procedures in manufacturing processes: recent studies and research perspectives, International Journal of Production Research

[30] Moustafa Gadalla & Deyi Xue, 2017, "An approach to identify the optimal configurations and reconfiguration processes for design of reconfigurable machine tools", Int. Journal of Production Research.

[31] Bortolini, M., Galizia, F.G. and Mora, C., 2018, "Reconfigurable Manufacturing Systems: Literature review and research trend," Journal of Manufacturing Systems, 49, pp. 93–106.

[32] Koren Y, Ulsoy AG., 1997, "Reconfigurable manufacturing systems. Engineering Research Center for Reconfigurable Machining Systems." ERC/RMS report #1. Ann Arbor

[33] Hu, S.J.,2013, "Evolving paradigms of manufacturing: From mass production to mass customization and Personalization" Procedia CIRP, 7

[34] Avinoam Tzimerman & Yale T. Herer, 2009, "Off-line inspections under inspection errors", IIE Transactions, 41:7, 626-641

[35] Bair N, Kidwai T, Koren Y, Mehrabi M, Wayne S, Prater L, 2002, "Design of a reconfigurable assembly system for manufacturing heat exchangers. Japan-USA Symposium on Flexible Manufacturing,

[36] Scholz, S., 2016, "A modular flexible scalable and reconfigurable system for manufacturing of Microsystems based on Additive Manufacturing and e-printing," Robotics and Computer-Integrated Manufacturing, 40, pp. 14–23

[37] Barhak J, Katz R (2003) Interpretation of laser measurements produced by the reconfigurable inspection machine using the "virtual ball" method. Proc CIRP- 2nd Intl. Conference on RMS '03, Ann Arbor, MI

[38] Tatyana I., Reich S., Bevec R., Goasar Z., Tamousinaite M., Ude A., Worgotter F., 2018, "Visual Inspection and Error Detection in a Reconfigurable Robot Workcell: An Automotive Light Assembly Example", Proceedings of the 13th International Joint Conference on Computer Vision, Imaging and Computer Graphics Theory and Applications